\documentstyle[amssymb,preprint,aps]{revtex}

\tightenlines

\begin{document}
\title{Polaritons in 2D-crystals and localized modes in narrow waveguides}
\author{V.S. Podolsky}
\address{Department of Physics, Queens College, CUNY, Flushing, NY 11367}
\date{\today}
\maketitle

\begin{abstract}
We study 2D-polaritons in an atomically thin dipole-active layer
(2D-crystal) placed inside a parallel-plate wavegude, and investigate the
possibility to obtain the localized wavegude modes associated with atomic
defects. Considering the wavegude width, $l,$ as an adjustable parameter, we
show that in the waveguide with $l\sim 10^{4}\,a,$\ where $a$ is the lattice
parameter, the localized mode can be provided by a single impurity or a
local structural defect.
\end{abstract}


\section{Introduction}

The mechanism of the photon localization {\it via} the resonant coupling
between photons and the local excitations inside the polariton gap was for
the first time proposed in Refs.\cite{Rupasov}. Considering a dipole-active
impurity atom in an isotropic frequency-dispersive medium, the authors
discovered the photon-atom bound states. They showed that when the atom
transition frequencies fall inside the polariton gap, the radiative
relaxation of the bound states is suppressed, and the field is localized
around the impurity.

In the recent papers \cite{Leva,Cubic}, we considered another type of the
impurity-induced polariton states associated with the phonon{\it \ }local
states in polar crystals. It was shown that near the bottom of the polariton
gap, the local states are predominated by the long-wavelength modes and have
the macroscopic localization and coherence radiuses. We attributed these
features to the singularity of the density of states at the gap bottom,
which also causes the absence of the lower threshold for the impurity
strength. The singularity is provided by the long-wavelength polaritons and
is generic for any isotropic dipole-active phonon mode with the negative
dispersion.

The long-wavelength nature of the polariton states allowed us to analyze the
crossover between the polariton and phonon local states within the continuum
approximation. Our results show that the crossover takes place in a
relativistically narrow interval near the bottom of the polariton gap \cite
{Crossover}. A small width of the crossover region and a strong suppression
of the photon content of the polariton local states are caused by the fact
that the typical momentum of the modes dominating in these states, $k_{\max
}\sim \beta ^{1/2}a^{-1},$ is much greater than the cross-resonance
momentum, $k_{0}\sim \beta a^{-1},$ where $a$ is the lattice parameter, and $%
\beta =v/c\,\ $is the ratio between the phonon velocity and the speed of
light. To eliminate this disproportion, one needs to lower the group
velocity of electromagnetic waves in the active medium. It can be achieved
if the medium is placed inside a waveguide. For instance, in the
parallel-plate waveguide, the dispersion law of the propagating modes, $%
\omega \!_{n}(k)=c\!\sqrt{(\pi n/2\,l)^{2}+k^{2}},$ provides the reduction
of their group velocity in the long-wavelength region. The phonon and photon
velocities become comparable in the cross-resonance region if the spacing
between plates, $l,$ is of the order of $10^{6}a$. However, \thinspace as
long as the waveguide spectrum contains the activationless mode $\left(
n=0\right) $, there is a guarantee that $k_{\max }$ is far away from the
cross-resonance point.

In the present paper, we investigate the polarization waves in an atomically
thin dielectric layer ( ionic 2D-crystal ) placed in a narrow parallel-plate
waveguide. The srtuctural stability of 2D-crystal is provided by a strong
confining potential, which also eliminates the activationless mode from the
waveguide spectrum. A sub-micron crystal film grown on a substrate can be a
physical realization of this model. The coherent interaction between the
waveguide modes and the polarization waves gives rise to new exitations -
2D-polaritons. They form two polariton bands, and the position of the
maximum \thinspace in the lower polariton branch depends on the waveguide
width. We show that $\,k_{\max }\sim k_{0}\,$\ for $l\sim \beta ^{-2/3}\,a,$
and $k_{\max }$ \thinspace tends to $\beta ^{2/3}\,a^{-1}$ when$\ l\gg \beta
^{-2/3}a$.\ In the latter case, the frequency region of the \thinspace
polariton local states is enlarged by the factor of $\beta ^{-1/3}\sim
10^{2},$ comparing to the 3D-case. The electric field in these states is
always delocalized across the 2D-crystal within the distance of the order of 
$l$ from the defect due to the contribution from the upper polariton band.
Upon increase of the impurity strength, the localization length decreases
and contributions to the field from the lower and upper branches begin to
compete everywhere in the waveguide. When the localization length becomes
comparable with the waveguide width, \thinspace the field is no longer
confined near the dielectric layer and the polariton local state transforms
into the localized waveguide mode. We estimate the corresponding value of
the impurity strength, and$\,$ show that, in the waveguide with $l\sim \beta
^{-2/3}\,a,$ the localized mode can be provided by a single impurity or a
structural defect.

\section{Polaritons in the ionic 2D-crystal}

Let us consider an atomically thin dielectric layer between two perfectly
conducting sheets [Fig.1]. The layer presents a regular ionic 2D-crystal,
which is stabilized in $z=0$ plane by a strong restraining potential. This
increases the activation energy of the off-layer lattice vibrations and
shifts the corresponding phonon modes much higher than the modes with the
in-plane polarization. Assuming an infinitely strong confining potential we
eliminate the off-plane phonons from our model. Among the remaining in-plane
phonons, for the sake of simplicity, we consider only a single transverse
dipole-active mode with the isotropic spectrum and negative dispersion.
\thinspace In the long-wavelength region, we can use the standard
approximation, $\Omega ^{2}(k)\approx \Omega _{0}^{2}-v^{2}k^{2},$ where $v$
sets the range of the typical phonon velocities. The dipole-active
excitations in a thin layer have a high decay rate, unless induced
electromagnetic radiation from the layer is compensated. In our model it is
provided by the coherent coupling between the lattice excitations and the
eigen modes of the waveguide. In the parallel plate waveguide, there are two
types of propagating modes [Fig.1]. TM-modes (transverse-magnetic) involve
electric field directed across the dielectric layer and, therefore, they
cannot be activated in our model. Electric field in TE-modes
(transverse-electric) is directed along the layer and, therefore, these
modes can be excited along with the the transverse-optical phonons.
\thinspace Introducing the surface polarization ${Q(}x{\bf ,}y{)}$
associated with these 2D-phonons and considering its dynamics together with
the corresponding TE-modes of the waveguide, we obtain the following system
of equations: 
\begin{equation}
c^{2}\frac{\partial ^{2}{E}\!_{\,\,{\bf k}}}{\partial z^{2}}+\left( \omega
^{2}-c^{2}k^{2}\!\right) \,{E}\!_{\,{\bf k}}=-4\pi \omega ^{2}{Q}_{{\bf k}%
}\!\,\delta \,(z),  \label{1}
\end{equation}

\begin{equation}
\lbrack \omega ^{2}-\Omega ^{2}(k)\,]\,{Q\,}\!_{{\bf k}}=-\frac{ad^{2}}{4\pi 
}\!{E}_{{\bf k}}{\bf (}0{\bf )},  \label{2}
\end{equation}
where ${Q\,}\!_{{\bf k}}$ and ${E}\!_{\,\,{\bf k}}(z)$ are the 2D-Fourier
amplitudes of the polarization and electric fields, ${\bf k}$ is a 2D-wave
vector, $d$ is the phonon-photon coupling parameter ( ion plasma frequency).
For the `` order of magnitude'' estimates we assume in this paper that $%
d\sim \Omega _{0}\sim v/a.$ For later convenience we introduce the
dimensionless variables $lk\rightarrow k,\,\,l\omega \,/\,c\rightarrow
\omega ,\ $and $l\Omega /c\rightarrow \Omega ,$ so that $\,\Omega
^{2}(k)\approx \Omega _{0}^{2}-\beta ^{2}k^{2},\,$and $\Omega _{0}\sim
k_{0}\sim \eta \beta ,\ \,$where $\eta =l/a$ is a large parameter.

The electric field in Eq.(1) is confined between the conducting plates and
its normal derivative at the dielectric sheet has a discontinuity caused by
the surface polarization. Solving Eqs.(1, 2) under these conditions, we
obtain the following dispersion equation: 
\begin{equation}
\frac{\Omega ^{2}-\omega ^{2}\,}{\omega ^{2}}=\delta \ \frac{\tan \sqrt{%
\omega ^{2}-k^{2}}}{\sqrt{\omega ^{2}-k^{2}}}\,.  \label{3}
\end{equation}
where $\delta =al\,d^{2}/\,2\,c^{2}\sim \eta \beta ^{2}$ is a small
parameter.

This equation defines a series of polariton branches, $\omega =\omega
_{n}\left( k\right) ,$ with their activation frequencies given by the
equation [Fig. 2,3]: 
\begin{equation}
\omega _{n}^{2}-\Omega _{0}^{2}=-\delta \omega _{n}\tan \omega _{n}\,.
\label{4}
\end{equation}
Analysis of Eq. (3) shows that the upper$\,\,\left( n\geq 1\right) \,\,$%
branches have the quadratic ``large momenta'' asymptotes: 
\begin{equation}
\omega _{n}^{2}\left( k\right) \thickapprox \omega _{n}^{2}+k^{2},  \label{5}
\end{equation}
where $\omega _{n}\approx \pi \left( n-1\,\,/\,2\right) \gg \Omega _{0}.$
Therefore, their spectral bands all overlap and form a common upper
polariton band.

The lower $\left( n=0\right) \,\,$polariton branch is separated from the
others by the polariton gap, the bottom of which coincides with the maximum
in this branch. Evaluation of the group velocity at the center of the
Brillouin zone, 
\begin{equation}
\left[ \frac{d\omega _{0}^{2}\left( k\right) }{dk^{2}}\right] _{k=0}\propto
-\,\beta ^{2}+\frac{\Omega _{0}^{2}-\,\omega _{0}^{2}}{2\omega _{0}^{2}}%
\left( \frac{2\omega _{0}}{\sin 2\omega _{0}}-1\right) \sim -\,\beta ^{2}+%
\frac{\delta \Omega _{0}^{2}}{3}\,\ ,  \label{6}
\end{equation}
shows that the lower branch has a negative dispersion if\thinspace
\thinspace \thinspace \thinspace $l\lesssim \beta ^{-2/3}a\sim 10^{4}a.\,\,$%
In such narrow waveguides the photon-phonon coupling is negligible and the
phonon branch, with its maximum located at $k=0,$ remains practically
unaffected by the field.

As $l$ increases, the maximum moves away from $k=0,$ reaching the
cross-resonance region, $k\sim k_{0},$ $\,$at $\,l\sim \beta ^{-2/3}a.$ We
restrict our further consideration to the case, \thinspace $l\gg \beta
^{-2/3}a,\,$only. It guarantees that the maximum of the lower branch is
located far away from $k_{0},$ in the region where $\,k\gg \Omega ,\omega ,$
and Eq.(3) can be approximated as follows: 
\begin{equation}
\Omega ^{2}-\omega ^{2}\approx \frac{\delta }{2k^{3}}\omega ^{2}\left(
\omega ^{2}+2k^{2}\right) \ .  \label{7}
\end{equation}
A positively defined solution of Eq.(7) gives the ``large momenta''
asymptote of the lower polariton branch, 
\begin{equation}
\omega _{0}^{2}(k)\approx \Omega ^{2}\left( 1-\frac{\delta }{k}\right) .
\label{8}
\end{equation}
The corresponding dispersion curve reaches its maximum at the point 
\begin{equation}
k_{\max }\approx l\left( \frac{ad^{2}\Omega _{0}^{2}}{2v^{2}c^{2}}\right)
^{1/3}\sim \eta \beta ^{2/3}\gg 1,  \label{9}
\end{equation}
where it sets the bottom of the polariton gap: 
\begin{equation}
\omega _{\max }^{2}\approx \Omega _{0}^{2}-3\frac{l^{2}}{c^{2}}\left( \frac{%
avd^{2}\Omega _{0}^{2}}{2c^{2}}\right) ^{2/3}\sim \Omega _{0}^{2}-3\beta
^{4/3}\left( d^{2}\Omega _{0}\right) ^{2/3}\,\frac{l^{2}}{c^{2}}\,\,.
\label{10}
\end{equation}
In the immediate vicinity of $k_{\max }$ the polariton dispersion law can be
presented as follows: 
\begin{equation}
\omega _{0}^{2}(k)\approx \omega _{\max }^{2}-3\beta ^{2}\left( k-k_{\max
}\right) ^{2},  \label{11}
\end{equation}
and the asymptote of the density of states near the gap bottom has the form: 
\begin{equation}
\rho (\omega ^{2})\propto \frac{k_{\max }}{2\pi \beta \sqrt{\,3\left( \omega
_{\max }^{2}-\omega ^{2}\right) }}\sim \frac{c}{a}\,\frac{\eta \beta ^{2/3}}{%
\Omega _{0}\,\sqrt{\omega _{\max }^{2}-\omega ^{2}}}.  \label{12}
\end{equation}
Restoring the true dimensionality of the variables and comparing our results
with those obtained for 3D-polaritons: 
\begin{eqnarray*}
k_{\max } &\sim &\beta ^{1/2}a^{-1},\ \omega _{\max }^{2}\sim \Omega
_{0}^{2}-\beta \,d\Omega _{0}, \\
\rho (\omega ^{2}) &\propto &\frac{a(ak_{\max })^{2}}{v\sqrt{\omega _{\max
}^{2}-\omega ^{2}}}\sim \frac{\beta }{\Omega _{0}\,\sqrt{\omega _{\max
}^{2}-\omega ^{2}}},
\end{eqnarray*}
one can see that the maximum of the polariton curve is now located closer to
the cross-resonance point, $k_{0}\sim \beta a^{-1},$ and the bottom of the
gap is shifted toward the phonon activation frequency. \thinspace Also, the
singularity of the density of states is strengthened by the factor of $%
\,\beta ^{-1/3}\sim 10^{2}.$ \thinspace Setting $\rho (\omega ^{2})\sim
1/\Delta ,\,$where $\Delta $ is the width of the polariton band, one can
estimate the frequency range where the singularity prevails: 
\begin{equation}
\sqrt{\omega _{\max }^{2}-\omega ^{2}}\sim \beta ^{2/3}\frac{\Delta }{\Omega
_{0}}\sim \beta ^{2/3}\Omega _{0}\,\,.  \label{13}
\end{equation}
Comparing to the 3D-case, it is enlarged by the factor of $\beta ^{-1/3}\sim
10^{2},$ what broadens\thinspace the region of the local states dominated by
long-wavelength modes.

\section{Defect-induced local states}

If a point-like defect is embedded in the dielectric layer, it modifies
Eq.(2) as follows: 
\begin{equation}
\lbrack \omega ^{2}-\Omega ^{2}(k)\,]\,{Q\,}\!_{{\bf k}}=-\frac{ad^{2}}{4\pi 
}\!{E}_{{\bf k}}{\bf (}0{\bf )+}\frac{\alpha a^{2}{Q}{\bf (0)}}{4\pi \,S},
\label{14}
\end{equation}
where ${Q}{\bf (0)\,\,}$is the polarization of a defect and $\,S\,$ is the
total area of the layer. The strength of the defect, $\alpha ,$ depends on
its mobility and binding energy in a crystal. In the case of an isotope
impurity $\alpha =-\omega ^{2}\delta m/m;$ for a non-isotope impurity or a
structural defect we assume that $\alpha \sim \overline{\alpha }\,\Omega
_{0}^{2},$ where $\overline{\alpha }$ is a numerical parameter.

Solving Eqs.(1,14) we obtain ( in the dimensionless variables $\omega ,k,$%
and $\Omega $): 
\begin{equation}
{E}_{\,{\bf k}}{\bf (}z{\bf )}=\frac{\alpha a^{2}l\,{Q}{\bf (0)}}{2\,S\,c^{2}%
}\times \frac{\omega ^{2}\sin \zeta {\cal Z}}{\zeta \,\cos \zeta }\times
\left( \omega ^{2}-\Omega ^{2}\,+\,\delta \,\frac{\,\omega ^{2}\,\tan \zeta 
}{\zeta }\right) ^{-1},  \label{15}
\end{equation}
where we denote $\,{\cal Z}=l^{-1}\left( l\,\pm z\right) \,\,\,$and $\zeta =%
\sqrt{\omega ^{2}-k^{2}}$.

$\,$Using this equation one can express ${Q}_{\,{\bf k}}$ {\it via} the
polarization of the defect, ${Q}{\bf (0),}$ and obtain then the spectral
equation for the local state: 
\begin{equation}
1=\frac{\alpha }{4\pi c^{2}}\left( \frac{a}{2\pi }\right) ^{2}\times \int d%
{\bf k}\left( \omega ^{2}-\Omega ^{2}+\,\delta \,\,\frac{\omega ^{2}\tan
\varkappa }{\varkappa }\right) ^{-1}.  \label{16}
\end{equation}
When the frequency approaches $\omega _{\max },$ the integral diverges at
the ``surface'' ${\bf k}^{2}=k_{\max }^{\,2}\,.$ Near $\omega _{\max }$ we
can approximate Eq.(16) as follows: 
\begin{equation}
1\approx \frac{\alpha }{2c^{2}}\left( \frac{a}{2\pi }\right) ^{2}\times
\int\limits_{0}^{\infty }\frac{kdk}{\omega ^{2}-\omega _{\max
}^{2}\,+\,3\beta ^{2}\,\left( k-k_{\max }\right) ^{2}},  \label{17}
\end{equation}
Retaining here only the singular part of the integral, we finally obtain: 
\begin{equation}
\sqrt{\omega ^{2}-\omega _{\max }^{2}}\approx \,\frac{\alpha a\left(
ak_{\max }\right) }{8\pi \,c^{2}\beta \sqrt{3}}\sim \overline{\alpha }\eta
\beta ^{5/3}.  \label{18}
\end{equation}
Comparing to the 3D-case $\left( \sqrt{\omega ^{2}-\omega _{\max }^{2}}\sim 
\overline{\alpha }\eta \beta ^{2}\right) ,$ the separation of the local
state from the bottom of the gap is enlarged by the factor of $\beta
^{-1/3}. $

Equation (18) defines the eigen frequency of the local state near the bottom
of the gap. To evaluate the localization radius of this state one needs to
consider the spatial distribution of the electric and the polarization
fields. The inverse Fourier transformation of Eq.\thinspace (15) gives us: 
\begin{equation}
{E}{\bf (}z,{\bf r)}=\frac{\alpha \,\omega ^{2}a^{2}\,\,{Q}{\bf (0)}}{4\pi
\,l\,c^{2}}\times \int\limits_{-\infty }^{\infty }\,\frac{\,dk\,k{\cal \,\,}{%
H}_{0}\left( {\cal R}k\right) \sin \left( {\cal Z}\zeta \right) }{\left(
\omega ^{2}-\Omega ^{2}\right) \zeta \cos \zeta +\delta \omega ^{2}\sin
\zeta }\,\,,  \label{19}
\end{equation}
where ${H}_{0}\left( {\cal R}k\right) \,\,\,$is the Hankel \thinspace
function of first kind, ${\cal R\,}=l^{-1}r\,$.

It follows from the spectral properties of our system, that the denominator
of the integrand, considered as a function of $\omega ^{2},$ has a series of
isolated simple zeroes: 
\begin{equation}
{F}\left( \omega ,k\right) =\left[ \left( \omega ^{2}-\Omega ^{2}\right)
\zeta \cos \zeta +\delta \omega ^{2}\sin \zeta \right] \varpropto
\prod_{n}\left[ \omega ^{2}-\omega _{n}^{2}\left( k\right) \right] \,\,,
\label{20}
\end{equation}
where the index $n$ enumerates different polariton branches, $\omega
_{n}^{2}\left( k\right) .$ Since the frequency of the local state, $\omega ,$
lies in the polatiton gap, all $k$-zeroes of ${F}\left( \omega ,k\right) $
are removed from the $%
\mathop{\rm Re}%
k$-axis. Closing the integration contour in Eq.(19) through the upper
half-plane, we can calculate the integral by the method of residuals: 
\begin{equation}
{E}{\bf (}z,{\bf r)}=\frac{i\alpha \,\omega ^{2}a^{2}\,{Q}{\bf (0)}}{%
2\,l\,c^{2}}\times \sum_{k_{n}}\,%
\mathop{\rm Res}%
\left\{ {H}_{0}\left( {\cal R}k\right) \sin \left( {\cal Z}\zeta \right)
\left[ {F}\left( \omega ,k\right) \right] ^{-1}\right\} _{k_{n}},  \label{21}
\end{equation}
where $k_{n}$ is a pole associated with $n$-th branch.

Near the bottom of the gap $\left( \omega \gtrsim \omega _{\max }\right) $
the factor presenting the lower polariton branch, $\omega ^{2}-\omega
_{0}^{2}\left( k\right) ,$ is small for $k\sim k_{\max }.$ \thinspace It
suggests that the poles of $\left[ {F}\left( \omega ,k\right) \right] ^{-1}$
corresponding to the lower branch are located in the vicinity of $k_{\max }.$
Using Eq.(11) one can find: 
\begin{equation}
k_{0}^{\pm }=k_{\max }\pm i\sqrt{\frac{\omega ^{2}-\omega _{\max }^{2}}{%
3\beta ^{2}}}=k_{\max }\pm i\varkappa .  \label{22}
\end{equation}
\thinspace Taking into account that  $k_{\max }\sim \eta \beta ^{2/3}\gg
\omega _{\max }\sim \eta \beta ,$ and$\ \varkappa \sim \overline{\alpha }%
\,\,k_{\max }\ll k_{\max }$ for a weak defect, $\,$one can obtain: 
\begin{equation}
\mathop{\rm Res}%
\left[ {F}\left( \omega ,k\right) \right] _{k_{0}^{+}}^{-1}=\left[ \frac{%
\partial {F}\left[ \omega _{0}\left( k\right) ,k\right] }{\partial k}\right]
_{k_{0}^{+}}^{-1}\approx \frac{\exp \left( -\,k_{\max }\right) }{3\beta
^{2}\varkappa \ k_{\max }}.  \label{23}
\end{equation}

\thinspace To evaluate zeroes of ${F}\left( \omega ,k\right) $ corresponding
to the upper branches, we first impose the upper limitation on the width of
the waveguide, such that $\beta ^{-2/3}a\ll l\ll $ $\beta ^{-1}a.$ In this
case, the phonon band lies well below the waveguide cut-off frequency and
all high-order polariton branches are practically indistinguishable from the
parent TE-modes. It gives us: 
\begin{equation}
\,\,k_{n}^{\pm }=\pm \,\,i\,\sqrt{\omega _{n}^{2}-\omega ^{2}}\thickapprox
\pm \,\,\,i\omega _{n}\,,  \label{24}
\end{equation}
and the corresponding residuals: 
\begin{equation}
\,%
\mathop{\rm Res}%
\left[ {F}\left( \omega ,k\right) \right] _{k_{n}^{+}}^{-1}\approx \frac{%
i\cos \omega _{n}}{\delta \ \omega ^{2}}.  \label{25}
\end{equation}
Finally, using Eqs.(21-26), we can find the large-distance asymptote of the
electric field: 
\begin{equation}
{E}{\bf (}z,{\bf r)}\varpropto {\cal R}^{-1/2}\left[ \frac{\exp \left( -%
{\cal R}\varkappa -zk_{\max }\right) }{6\,\,\beta ^{2}\,\varkappa
\,\,k_{\max }^{\,3/2}}+\sum_{n=1}^{\infty }\frac{\exp \left( -{\cal R}\omega
_{n}\right) \sin \left( {\cal Z}\omega _{n}\right) \cos \omega _{n}}{\delta
\ \omega ^{2}\ \omega _{n}^{1/2}}\right] .  \label{26}
\end{equation}

This result presents the field as a sum of contributions from all branches.
However, because $\omega _{n}\ $lie close to $\pi \left( n-\frac{1}{2}%
\right) $ for the upper branches, the only several  terms corresponding to
the low-located branches can be retained in the sum.

Equation (26) explicitely demonstrates the localization of the field in
radial directions, whereas, the$\ z$-profile of the field depends on the
defect strength, $\overline{\alpha }.$ For a weak defect, when $\varkappa
\ll \omega _{1}$(\thinspace localization length is much greater than $l\,$),
the first term of Eq. (26) dominates at ${\cal R}\gg \varkappa ^{-1}$ and
confines the field within the $k_{\max }^{-1}$ -wide layer around the
dielectric sheet. However, in the kernel of the local state, within the
distance ${\cal R}\lesssim \varkappa ^{-1}$ from the defect, next terms of
Eq.(26) begin to compete with the lower band contribution. This leads to
``delocalization'' of the\thinspace field in $z$-direction within the kernel
.

Upon increase of the defect strength, the local state moves away from $%
\omega _{\max }$ and the localization length, $\varkappa ^{-1},$decreases.
\thinspace When $\,$we are reaching $\overline{\alpha }\,_{%
\mathop{\rm cr}%
}\sim \omega _{1}/k_{\max },$ the first term of Eq.(25) is no longer
dominating in ${E}{\bf (}z,{\bf r).}$ \thinspace In this case, despite the
localization of the field within the $l$-range in radial directions, its $z$
-confinement completely disappears. Such a state can be qualified as a local
waveguide mode. Analysis of the structure and properties of these states
\thinspace requires one to consider waveguides with $l\sim \beta ^{-2/3}a,$
where $k_{\max }\sim k_{0}.$ Recalling the result concerning  the energy
distribution for 3D local polaritons  \cite{Crossover}, $W_{{\rm field}}/W_{%
{\rm mech}}\sim \left( \,k_{0}/k_{\max }\right) ^{4},$ we can expect the
energy equipartition in the local waveguide modes. Also, since $k_{\max
}\sim \omega _{1}$ in the sufficiently narrow waveguides, there the local
modes can be provided by defects with $\overline{\alpha }\lesssim 1,$ such
as isotop impurities or local structural defects.  A more detailed analysis
of this case will be presented elsewhere.

\end{document}